\documentstyle[twoside,psfig]{article}

\catcode`\@=11
\long\def\@makefntext#1{
\protect\noindent \hbox to 3.2pt {\hskip-.9pt  
$^{{\eightrm\@thefnmark}}$\hfil}#1\hfill}		%CAN BE USED 

\def\@makefnmark{\hbox to 0pt{$^{\@thefnmark}$\hss}}	%ORIGINAL 
	
\def\ps@myheadings{\let\@mkboth\@gobbletwo
\def\@oddhead{\hbox{}
\rightmark\hfil\eightrm\thepage}   
\def\@oddfoot{}\def\@evenhead{\eightrm\thepage\hfil
\leftmark\hbox{}}\def\@evenfoot{}
\def\sectionmark##1{}\def\subsectionmark##1{}}

\oddsidemargin=\evensidemargin
\newcounter{sectionc}\newcounter{subsectionc}\newcounter{subsubsectionc}
\renewcommand{\section}[1] {\vspace{12pt}\addtocounter{sectionc}{1} 
\setcounter{subsectionc}{0}\setcounter{subsubsectionc}{0}\noindent 
	{\tenbf\thesectionc. #1}\par\vspace{5pt}}
\renewcommand{\subsection}[1] {\vspace{12pt}\addtocounter{subsectionc}{1} 
	\setcounter{subsubsectionc}{0}\noindent 
	{\bf\thesectionc.\thesubsectionc. {\kern1pt \bfit #1}}\par\vspace{5pt}}
\renewcommand{\subsubsection}[1] {\vspace{12pt}\addtocounter{subsubsectionc}{1}
	\noindent{\tenrm\thesectionc.\thesubsectionc.\thesubsubsectionc.
	{\kern1pt \tenit #1}}\par\vspace{5pt}}

\newcounter{appendixc}
\newcounter{subappendixc}[appendixc]
\newcounter{subsubappendixc}[subappendixc]
\renewcommand{\thesubappendixc}{\Alph{appendixc}.\arabic{subappendixc}}
\renewcommand{\thesubsubappendixc}
	{\Alph{appendixc}.\arabic{subappendixc}.\arabic{subsubappendixc}}

\renewcommand{\appendix}[1] {\vspace{12pt}
        \refstepcounter{appendixc}
        \setcounter{figure}{0}
        \setcounter{table}{0}
        \setcounter{lemma}{0}
        \setcounter{theorem}{0}
        \setcounter{corollary}{0}
        \setcounter{definition}{0}
        \setcounter{equation}{0}
        \renewcommand{\thefigure}{\Alph{appendixc}.\arabic{figure}}
        \renewcommand{\thetable}{\Alph{appendixc}.\arabic{table}}
        \renewcommand{\theappendixc}{\Alph{appendixc}}
        \renewcommand{\thelemma}{\Alph{appendixc}.\arabic{lemma}}
        \renewcommand{\thetheorem}{\Alph{appendixc}.\arabic{theorem}}
        \renewcommand{\thedefinition}{\Alph{appendixc}.\arabic{definition}}
        \renewcommand{\thecorollary}{\Alph{appendixc}.\arabic{corollary}}
        \renewcommand{\theequation}{\Alph{appendixc}.\arabic{equation}}
        \noindent{\tenbf Appendix \theappendixc #1}\par\vspace{5pt}}
\newcommand{\subappendix}[1] {\vspace{12pt}
        \refstepcounter{subappendixc}
        \noindent{\bf Appendix \thesubappendixc. {\kern1pt \bfit #1}}
	\par\vspace{5pt}}
\newcommand{\subsubappendix}[1] {\vspace{12pt}
        \refstepcounter{subsubappendixc}
        \noindent{\rm Appendix \thesubsubappendixc. {\kern1pt \tenit #1}}
	\par\vspace{5pt}}

\newcommand{\textlineskip}{\baselineskip=13pt}
\newcommand{\smalllineskip}{\baselineskip=10pt}

\def\eightcirc{
\begin{picture}(0,0)
\put(4.4,1.8){\circle{6.5}}
\end{picture}}
\def\eightcopyright{\eightcirc\kern2.7pt\hbox{\eightrm c}} 

\newcommand{\copyrightheading}[1]
	{\vspace*{-2.5cm}\smalllineskip{\flushleft
	{\footnotesize Modern Physics Letters A #1}\\
	{\footnotesize $\eightcopyright$\, World Scientific Publishing
	 Company}\\
	 }}

\renewenvironment{thebibliography}[1]
	{\frenchspacing
	 \ninerm\baselineskip=11pt
	 \begin{list}{\arabic{enumi}.}
        {\usecounter{enumi}\setlength{\parsep}{0pt}     
	 \setlength{\leftmargin 12.7pt}{\rightmargin 0pt} %FOR 1--9 ITEMS
         \setlength{\itemsep}{0pt} \settowidth
	{\labelwidth}{#1.}\sloppy}}{\end{list}}

\newcounter{itemlistc}
\newcounter{romanlistc}
\newcounter{alphlistc}
\newcounter{arabiclistc}

\newcommand{\fcaption}[1]{
        \refstepcounter{figure}
        \setbox\@tempboxa = \hbox{\footnotesize Fig.~\thefigure. #1}
        \ifdim \wd\@tempboxa > 5in
           {\begin{center}
        \parbox{5in}{\footnotesize\smalllineskip Fig.~\thefigure. #1}
            \end{center}}
        \else
             {\begin{center}
             {\footnotesize Fig.~\thefigure. #1}
              \end{center}}
        \fi}

\newcommand{\tcaption}[1]{
        \refstepcounter{table}
        \setbox\@tempboxa = \hbox{\footnotesize Table~\thetable. #1}
        \ifdim \wd\@tempboxa > 5in
           {\begin{center}
        \parbox{5in}{\footnotesize\smalllineskip Table~\thetable. #1}
            \end{center}}
        \else
             {\begin{center}
             {\footnotesize Table~\thetable. #1}
              \end{center}}
        \fi}

\def\@citex[#1]#2{\if@filesw\immediate\write\@auxout
	{\string\citation{#2}}\fi
\def\@citea{}\@cite{\@for\@citeb:=#2\do
	{\@citea\def\@citea{,}\@ifundefined
	{b@\@citeb}{{\bf ?}\@warning
	{Citation `\@citeb' on page \thepage \space undefined}}
	{\csname b@\@citeb\endcsname}}}{#1}}

\newif\if@cghi
\def\cite{\@cghitrue\@ifnextchar [{\@tempswatrue
	\@citex}{\@tempswafalse\@citex[]}}
\def\citelow{\@cghifalse\@ifnextchar [{\@tempswatrue
	\@citex}{\@tempswafalse\@citex[]}}
\def\@cite#1#2{{$\null^{#1}$\if@tempswa\typeout
	{IJCGA warning: optional citation argument 
	ignored: `#2'} \fi}}

\def\pmb#1{\setbox0=\hbox{#1}
	\kern-.025em\copy0\kern-\wd0
	\kern.05em\copy0\kern-\wd0
	\kern-.025em\raise.0433em\box0}

\def\fnt#1#2{\footnotetext{\kern-.3em
	{$^{\mbox{\scriptsize #1}}$}{#2}}}

\def\ps@myheadings{%
    \let\@oddfoot\@empty\let\@evenfoot\@empty
    \def\@evenhead{\slshape\leftmark\hfil}%       %EVEN PAGE
    \def\@oddhead{\hfil{\slshape\rightmark}}%     %ODD PAGE
    \let\@mkboth\@gobbletwo
    \let\sectionmark\@gobble
    \let\subsectionmark\@gobble
    }
\font\tenrm=cmr10
\font\tenit=cmti10 
\font\tenbf=cmbx10
\font\bfit=cmbxti10 at 10pt
\font\ninerm=cmr9

\font\eightrm=cmr8

\def\qed{\hbox{${\vcenter{\vbox{			%HOLLOW SQUARE
   \hrule height 0.4pt\hbox{\vrule width 0.4pt height 6pt
   \kern5pt\vrule width 0.4pt}\hrule height 0.4pt}}}$}}

\begin{document}
\setlength{\textheight}{7.7truein}  %for 2nd page onwards

\thispagestyle{empty}

%\markboth{\protect{\footnotesize\it Instructions for Typesetting
%Manuscripts}}{\protect{\footnotesize\it Instructions for
%Typesetting Manuscripts}}

\normalsize\textlineskip

\setcounter{page}{1}

\copyrightheading{}	%{Vol.~0, No.~0 (2002) 000--000}

\vspace*{0.88truein}

\baselineskip=13pt
\centerline{\bf  SPIN RESPONSE OF THE NUCLEON}
\centerline{\bf IN THE RESONANCE REGION\footnote{Talk presented 
at the Second Asia Pacific 
Conference on Few Body Problems in Physics, Shanghai, 
China, August 27 - 30, 2002}}
\baselineskip=13pt
%\vspace*{0.37truein}
\vspace*{0.4truein}
\centerline{\footnotesize VOLKER D. BURKERT}
\baselineskip=12pt
\centerline{\footnotesize\it Jefferson Lab, 12000 Jefferson Avenue}
\baselineskip=10pt
\centerline{\footnotesize\it Newport News, Virgnia 23606, USA}
%\vspace*{10pt}
\vspace*{12pt}

%\publisher{(received date)}{(revised date)}

%\vspace*{0.21truein}
\vspace*{0.23truein}
\begin{abstract} {I discuss recent results from CLAS and Hall A at Jefferson Lab on 
the measurement 
of inclusive spin structure functions in the nucleon resonance region 
using polarized electron beams and polarized targets.
Results on the first moment of the spin structure function for 
protons and neutrons will be discussed, as well as the Bjorken integral. 
I will argue that the helicity structure of individual 
resonances plays a vital role in understanding the nucleon's spin response
in the domain of strong interaction QCD, and must be considered in 
any analysis of the nucleon spin structure at low and intermediate 
photon virtuality.}

\end{abstract}
%\vspace*{10pt}
%\keywords{The contents of the keywords}

\vspace*{2pt}

%\textlineskip			%) USE THIS MEASUREMENT WHEN THERE IS
%\vspace*{12pt}			%) NO SECTION HEADING

\baselineskip=13pt	        %) ACTUAL LEADING
\normalsize              	%) USE THIS MEASUREMENT WHEN THERE IS

\section{Introduction}
For more than 20 years, measurements of polarized structure functions in 
lepton 
nucleon scattering have been a focus of nucleon structure physics at
high energy laboratories. One of the surprising findings of the EMC 
experiment at CERN was that only a small fraction  of the nucleon spin is 
accounted for by the spin of the quarks\cite{ashman}. The initial results were 
confirmed by several follow-up experiments\cite{filipone}. This result is in 
conflict with simple quark model expectations, and demonstrated that we are far 
from having a realistic picture of the nucleon's internal structure. 
These experiments also studied the fundamental Bjorken sum 
rule\cite{bjorken} which,
at asymptotic momentum transfer, relates the proton-neutron difference 
of the first moment $\Gamma_1 = \int{g_1(x)dx}$ to 
the weak axial coupling constant: $\Gamma_1^p - \Gamma_1^n = {1\over 6}g_A~.$  
This sum rule has been evolved to the finite 
$Q^2$ values reached in the experiments using pQCD, and has been verified 
at the 5\% level.

While these measurements were carried out in the deep inelastic regime, 
the nucleon's spin response has hardly been 
measured in the low $Q^2$ regime and in the domain of nucleon resonances, 
which is the true domain of strong QCD. Our understanding of nucleon 
structure is incomplete, at best, if the nucleon is not also probed 
and fundamentally described at medium and large distance scales. 
This is the domain where current experiments at JLab have their 
greatest impact.

While the Bjorken sum rule provides a fundamental constraint at large $Q^2$, the 
Gerasimov-Drell-Hearn (GDH) sum rule \cite{gerasimov,drell} constrains the 
evolution at very low $Q^2$. The GDH sum rule 
relates the differences in the helicity-dependent total photoabsorption 
cross sections to the anomalous magnetic moment $\kappa$ of the target 
$${M^2 \over 8\pi^2\alpha}\int_{\nu_0}^{\infty}
{{{\sigma_{1/2}(\nu)-\sigma_{3/2}(\nu)}\over \nu}d\nu =
 -{1\over 4}\kappa^2}\eqno(1)$$  
\noindent
where $\nu_0$ is the photon energy at pion threshold, and M is the nucleon 
mass. The GDH sum rule also defines the slope of $\Gamma_1(Q^2=0)$, where 
the elastic contribution at $x=1$ has been excluded: 
$$2M^2{d\Gamma_1 \over dQ^2}(Q^2\rightarrow 0) = 
-{1\over 4}\kappa^2 \eqno(2)$$  
The sum rule has been studied for photon energies up
to 2.5 GeV \cite{ahrens}, and in this limited energy range deviates from 
the theoretical asymptotic value by less than 10\%.   
A rigorous extension of the sum rule to finite $Q^2$ has  
been introduced by Ji and Osborne \cite{jios}. 
Measurement of the $Q^2$-dependence of (1) allows tests of the low energy QCD
predictions of the GDH sum rule evolution in ChPT, and shed light on the question
at what distance scale pQCD corrections and the QCD twist 
expansion will break down, and where the physics of confinement will dominate. 
It will also allow us to evaluate where resonances give important 
contributions to the first moment\cite{burli,buriof}, as well 
as to the higher $x$-moments of 
the spin structure function $g_1(x,Q^2)$. 
The moments need to be determined experimentally and calculated 
in QCD. 
The well known ``duality'' between the deep inelastic regime, and the 
resonance regime observed for the unpolarized structure function $F_1(x,Q^2)$, 
needs to be 
explored for the spin structure function $g_1(x,Q^2)$. This will shed new 
light on this phenomenon.          
The first round of experiments has been completed on polarized
hydrogen ($NH_3$), deuterium ($ND_3$), and on $^3He$. On the theoretical side
we now see the first full (unquenched) QCD calculations for the electromagnetic 
transition from the ground state nucleon to 
the $\Delta(1232)$~\cite{alexandrou}. 
Results for other states, and coverage of a larger $Q^2$ range may 
soon be available. This may provide the basis for a future QCD description of 
the helicity structure of prominent resonance transitions.

\section{Expectations for  $\Gamma_1(Q^2)$}

The inclusive doubly polarized cross section can be written as:
$${1\over \Gamma_T} {d\sigma \over d\Omega dE^{\prime}} = \sigma_T 
+ \epsilon\sigma_L + P_eP_t[\sqrt{1-\epsilon^2}A_1\sigma_T\cos{\psi} + 
\sqrt{2\epsilon(1+\epsilon)}A_2\sigma_T\sin{\psi}]~\eqno(3)$$
where $A_1$ and $A_2$ are the spin-dependent asymmetries, $\psi$ is the angle between the nucleon polarization vector and the $\vec q$ vector, $\epsilon$ the
polarization parameter of the virtual photon, and $\sigma_T$ and $\sigma_L$ 
are the total absorption cross sections for transverse and longitudinal 
virtual photons. For electrons and nucleons polarized along the beam line, the 
experimental double polarization asymmetry is given by 
$$A_{exp} = P_eP_tD {A_1+\eta A_2 \over 1+\epsilon R}\eqno(4)$$
where D and $\eta$ are kinematic factors, $\epsilon$ describes the polarization of 
the virtual photon, and $R = \sigma_L/\sigma_T$.  
The asymmetries $A_1$ and $A_2$ are related to the spin structure function $g_1$ 
by  $$g_1(x,Q^2) = {\tau \over 1+\tau}
[A_1 + {1\over \sqrt{\tau}}A_2]F_1(x,Q^2) \eqno(5)$$ 
where $F_1$ is the unpolarized structure function, and $\tau = \nu^2/Q^2$.

The GDH and Bjorken sum rules provide constraints at the kinematic 
endpoints $Q^2=0$ and $Q^2 \rightarrow \infty$. 
The evolution of the Bjorken sum rule to finite values of $Q^2$ using 
pQCD and the twist expansion allow to connect experimental values measured at
finite $Q^2$ to the endpoints. 
Heavy Baryon Chiral Perturbation Theory (HBChPT) has been 
proposed as a tool to evolve the GDH sum rule to $Q^2 > 0$, possibly to
$Q^2 = 0.1$ GeV$^2$, and to use the twist expansion down to $Q^2=0.5$ GeV$^2$ 
\cite{ji}. 
If this is a realistic concept, and if lattice QCD can be used 
to describe prominent resonance contributions to $\Gamma_1(Q^2)$ in the 
range $Q^2 = 0.1 - 0.5$ GeV$^2$, {\sl this could provide the basis for a 
description 
of a basic quantity of nucleon structure physics from small to large 
distances within fundamental theory}, a worthwhile goal! 
 
Using the constraints given by the two endpoint sum rules we may already
get a qualitative picture of $\Gamma^p_1(Q^2)$ and $\Gamma^n_1(Q^2)$. 
There is no sum rule for the proton and neutron separately that
has been verified. However, experiments have determined the asymptotic 
limit with sufficient confidence for the proton and the neutron. At large 
$Q^2$, $\Gamma_1$  is expected to approach this limit following the 
pQCD evolution from finite values of $Q^2$. At small $Q^2$, $\Gamma_1$ 
must approach zero with a slope given by (2). 

Heavy Baryon ChPT in the lowest non trivial order predicts \cite{xji} 
$$2M_p^2  \Gamma_1^p(Q^2) = - {\kappa_p^2 \over 4}Q^2 + 6.85Q^4 (GeV^2) + .~.~.~\eqno(6)$$

\vspace{-0.5cm}
$$2M_n^2  \Gamma_1^n(Q^2) = - {\kappa_n^2 \over 4}Q^2 + 5.54Q^4 (GeV^2) + .~.~.~\eqno(7)$$

Unfortunately, the large coefficients of the $Q^4$ terms make the convergence 
of this
expansion unlikely for $Q^2 > 0.1$~GeV$^2$.
However, for the proton-neutron difference the situation is quite 
different\cite{burk} 
$$2M^2  \Gamma_1^{(p-n)}(Q^2) = - {{\kappa_p^2 - \kappa_n^2} \over 4}Q^2 + 1.31Q^4 (GeV^2) + .~.~.\eqno(8)$$
The $Q^4$ coefficient is a factor 4-5 smaller than for the proton or neutron, and one might
expect convergence up to considerably higher $Q^2$ than for proton and neutron separately. 
This may be due to
the absence of the $\Delta(1232)$ in $\Gamma_1^{(p-n)}$, and may hint at 
difficulties in describing the  $\Delta(1232)$ contributions in HBChPT. 
 
\clearpage

%%%%%%%%%%%%%%%%%%%%%%%%%%%%%%%%%%%%%%%%%%%%%%%%%%%
\begin{figure}[tb]
\vspace{54mm} 
\centering{\includegraphics{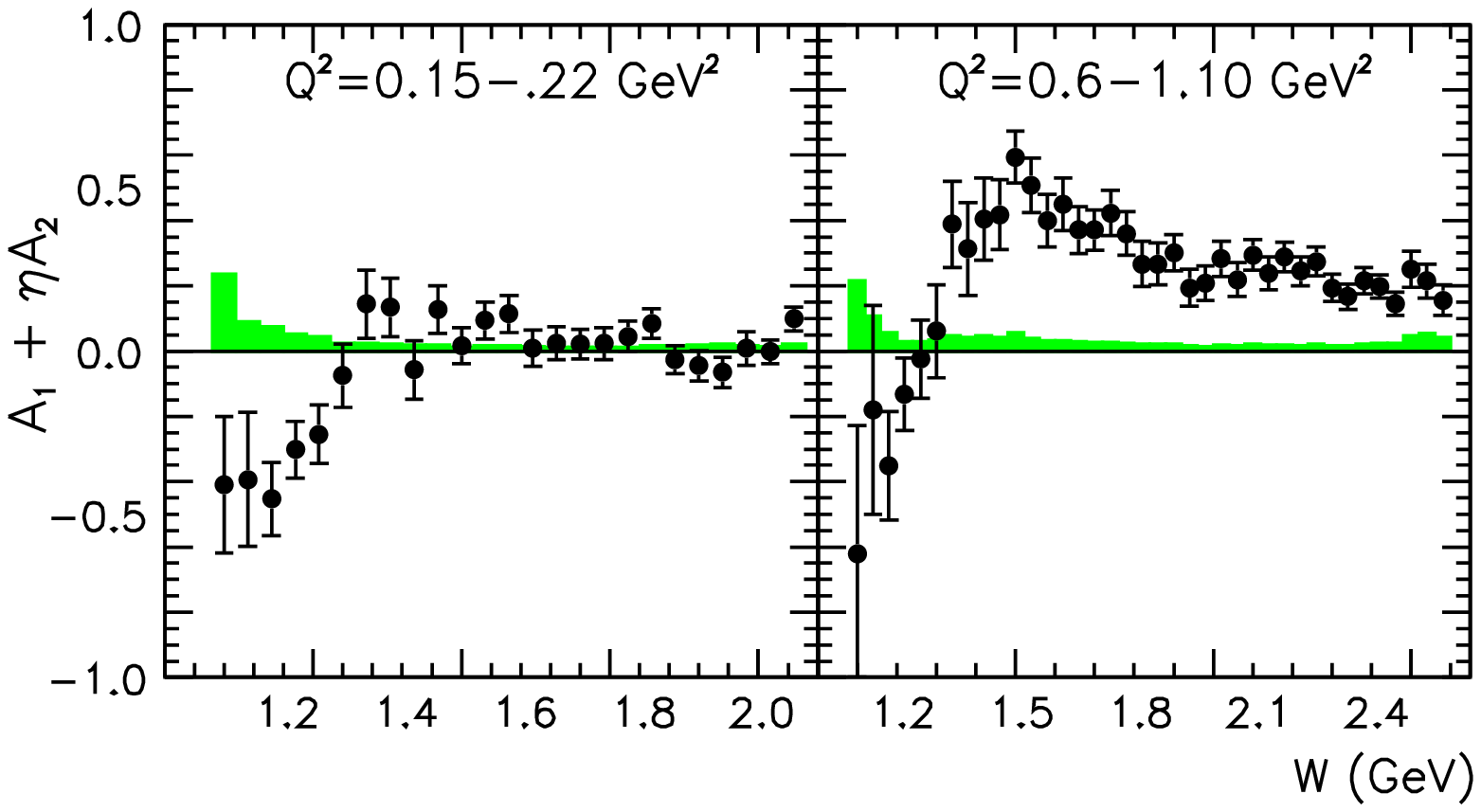}}
\caption{\small Asymmetry $A_1+\eta A_2$ for protons. The 
panels show results for two $Q^2$ values measured at 2.6 and 4.3 GeV beam energies.}
\label{fig:epasym}
%\end{figure}
%%%%%%%%%%%%%%%%%%%%%%%%%%%%%%%%%%%%%%%%%%%%%%%%%%%%%
%%%%%%%%%%%%%%%%%%%%%%%%%%%%%%%%%%%%%%%%%%%%%%%%%%%
%\begin{figure}[tb]
\vspace{60mm} 
\centering{\includegraphics{d13_asym.epsi}}
\caption{\small Helicity asymmetry $A_1(Q^2)$ for the $D_{13}(1520)$ 
resonance transition. $A_1$ has been extracted from partial wave analyses of 
single pion electroproduction  data.}
\label{fig:d13}
\end{figure}
%%%%%%%%%%%%%%%%%%%%%%%%%%%%%%%%%%%%%%%%%%%%%%%%%%%%%

\section{Results for Protons and Neutrons.}

Inclusive double polarization experiments have been carried out for energies 
of 2.6 and 4.3 GeV using $N\vec {H_3}$ \cite{bucramin} as polarized 
hydrogen target with CLAS. After subtracting the nuclear background measured in 
separate data runs, and using a parameterization of previous unpolarized 
measurements for $R$, equation (4) is used to
determine the asymmetry  $A_1+\eta A_1$. More details of the analysis can be found in
ref.\cite{minehart}. Two of several $Q^2$ bins are shown in Figure
\ref{fig:epasym}. 
In the lowest $Q^2$ bin, the asymmetry is dominated by 
the excitation of the $\Delta(1232)$ giving a significant negative contribution to 
$A_1$. At higher $Q^2$ the asymmetry in the
$\Delta(1232)$ region remains negative, but at higher W the asymmetry 
quickly becomes positive and large, reaching peak values of about 
0.6 at $Q^2=0.8$ GeV$^2$ and W=1.5 GeV. Evaluations of resonance
contributions show that this is largely driven by the 
$S_{11}(1535)$ $A_{1/2}$ amplitude, and by the rapidly changing helicity 
structure of the $D_{13}(1520)$ state. The latter resonance is 
known to have a dominant $A_{3/2}$ amplitude at the photon point, but 
is rapidly changing to $A_{1/2}$ dominance for $Q^2 > 0.5$ GeV$^2$. 
The helicity asymmetry $A_1(D_{13}(1520))$ is shown in Figure \ref{fig:d13}.

Using a parameterization of the world data on $F_1(x,Q^2)$ and 
$A_2(x,Q^2)$, we can extract $g_1(x,Q^2)$ from (5). Results 
are shown in Figure \ref{fig:g1x}. 
The main feature at low $Q^2$ is due to the negative contribution of the
$\Delta(1232)$ resonance. With increasing $Q^2$, however, the absolute strength of the $\Delta(1232)$ contribution decreases, while contributions of higher 
mass resonances increase and become more positive. Note, that higher mass 
contributions at fixed $Q^2$ appear at lower $x$ in this graph. 
The graphs also show a model parameterization 
of $g_1(x,Q^2)$ which is used to extrapolate to $x \rightarrow 0$. The model 
is based on a parametrization of the resonance transition formfactors and 
also describes the behavior of the spin structure functions in the 
deep inelastic regime.

\subsection{Is Quark-Hadron Duality valid for $g_1$ of the Proton?}

\begin{figure}[tb]
\vspace{75mm} 
\centering{\includegraphics{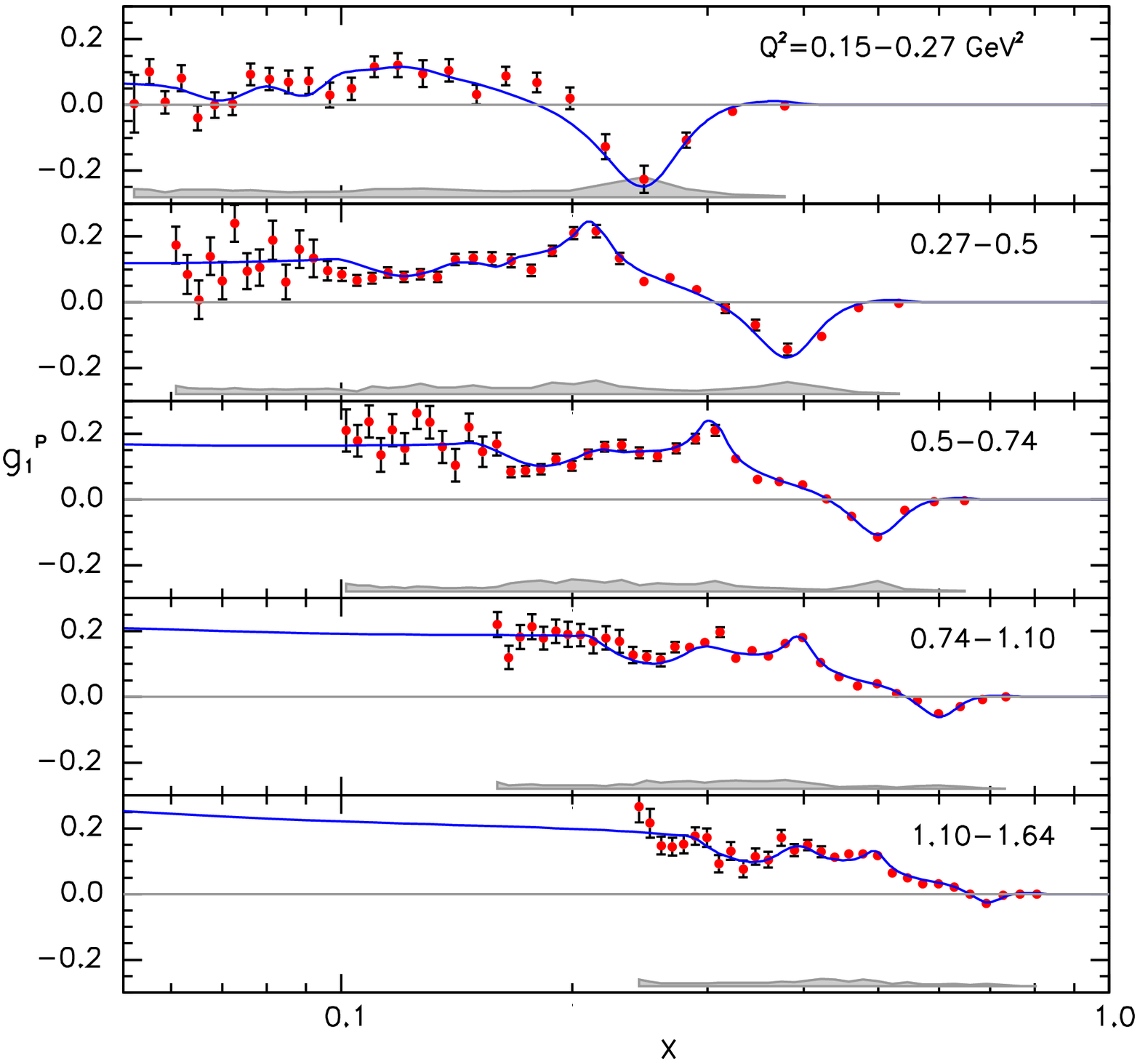}}
\caption{\small Spin structure function $g_1(x,Q^2)$ for the 
proton. The curve is a parameterization tuned to fit the JLab data, and 
is linked to the deep inelastic region based on prior knowledge. It is used 
for radiative corrections, and to extrapolate $g_1$ to $x=0$ for evaluation 
of $\Gamma_1$.}
\label{fig:g1x}
%\end{figure}
%%%%%%%%%%%%%%%%%%%%%%%%%%%%%%%%%%%%%%%%%%%%%%%%%%%%%   
%%%%%%%%%%%%%%%%%%%%%%%%%%%%%%%%%%%%%%%%%%%%%%%%%%%
%\begin{figure}[tb]
\vspace{85mm} 
\centering{\includegraphics{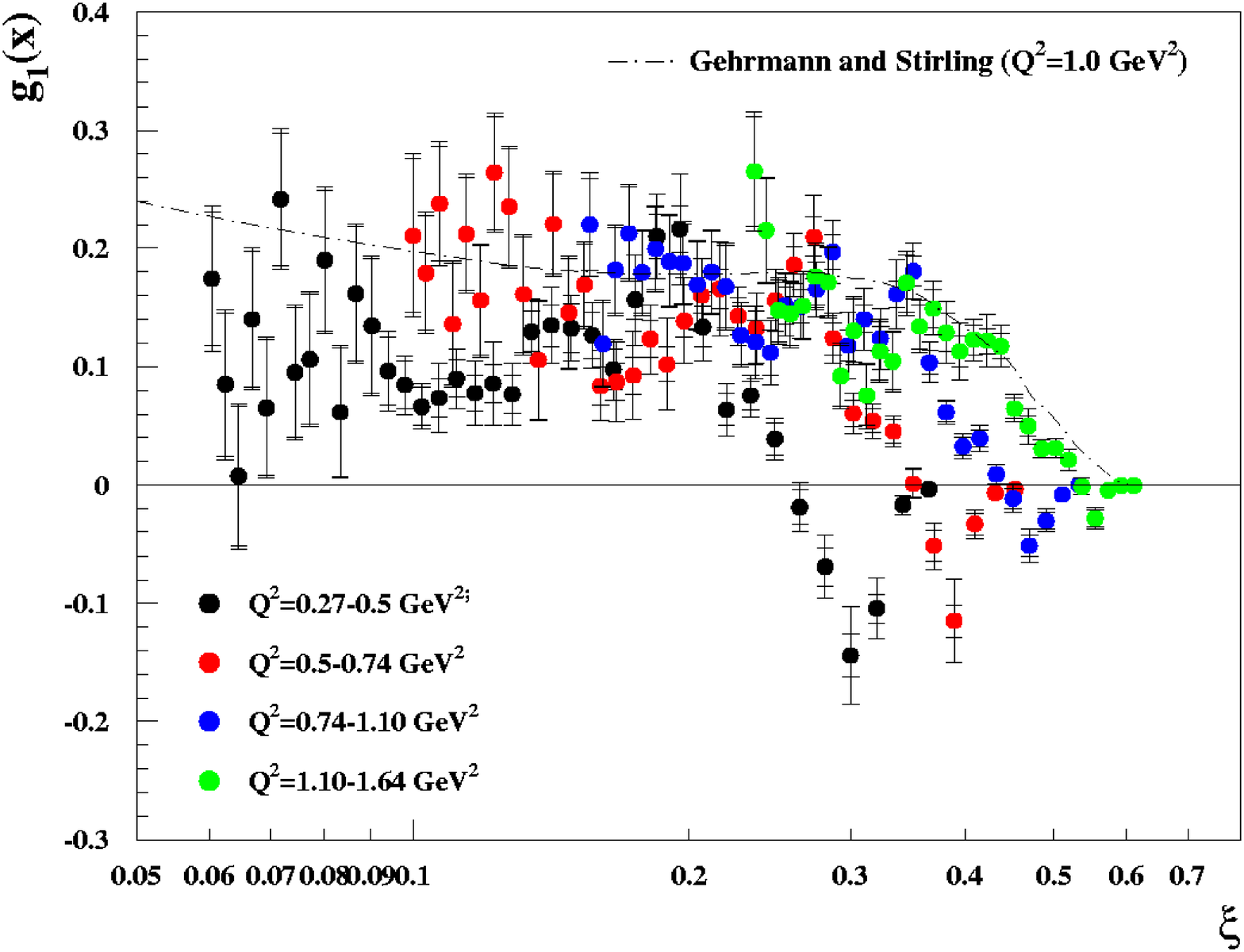}}
\caption{\small Duality test for $g_1$ on the proton. The Nachtmann 
scaling variable $\xi$ is used to account for target mass effects.}
\label{fig:duality}
\end{figure}

More than 3 decades ago, Bloom and Gilman\cite{bloom} found that parametrizations 
of inclusive unpolarized structure functions, measured in the deeply 
inelastic regime, approximately describe the resonance region provided one
averages over the resonance bumps. This phenomenon is known as local duality.
By comparing $g_1$ at various $Q^2$ we can infer
if such a behavior is also observed for the polarized structure function.
For the relatively low $Q^2$ measured in this experiment the Nachtmann 
variable  $\xi = 2x/(1+\sqrt{1+4x^2m^2/Q^2})$, which accounts for target 
mass effects, is a more 
appropriate scaling variable than the Bjorken variable $x$. 
Figure \ref{fig:duality} shows $g_1(\xi, Q^2)$ for the proton in comparison
with the scaling curve describing the deeply inelastic behavior. 

The negative contribution of the $\Delta(1232)$ obviously prevents a 
naive ``local duality'' to work for $Q^2 < 1.1$ GeV$^2$. 
Recently, Close and Isgur discussed in a simple harmonic oscillator 
model\cite{cloisg} that local duality is expected to work only if one 
integrates 
over states belonging to certain multiplets within the $SU(6)$ symmetry
group. In this case, for local duality to work for the $\Delta(1232)$, 
one would also need to include contributions from the proton ground state,
which belongs to the same multiplet $[56,0^+]$ as the $\Delta(1232)$. 
The positive contribution of elastic scattering to $g_1$ could therefore
offset the negative $\Delta$ contribution. Detailed duality tests for the 
higher mass resonances will
require a separation of overlapping states belonging to the same multiplet,
and measurement of their transition amplitudes. Such a program is currently
underway at JLab\cite{burkert_elba}.

%%%%%%%%%%%%%%%%%%%%%%%%%%%%%%%%%%%%%%%%%%%%%%%%%%%
%%%%%%%%%%%%%%%%%%%%%%%%%%%%%%%%%%%%%%%%%%%%%%%%%%%%%
\clearpage

%%%%%%%%%%%%%%%%%%%%%%%%%%%%%%%%%%%%%%%%%%%%%%%%%%%
\begin{figure}[tb]
\vspace{85mm} 
\centering{\includegraphics{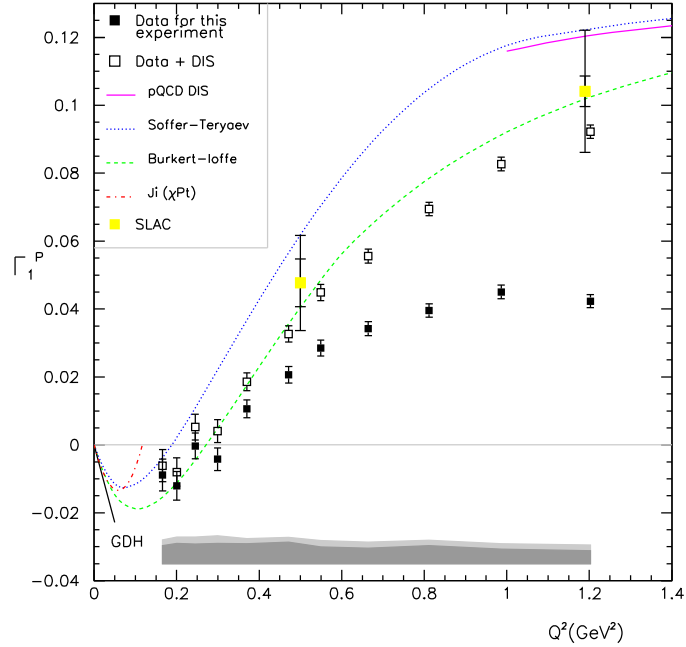}}
\caption{\small First moment $\Gamma_1(Q^2)$ for the proton. 
The black symbols correspond to the measured values from CLAS. 
The open squares are CLAS data 
corrected for the DIS contributions. Data from SLAC are shown for 
comparison.}
\label{fig:gammaboth}
\end{figure}
%%%%%%%%%%%%%%%%%%%%%%%%%%%%%%%%%%%%%%%%%%%%%%%%%%%%%

\section{The First Moment of Structure Function $g_1$}      

In order to obtain the first moment, the integral $\int^{1}_0 g_1(x, Q^2)dx $ 
is computed using the measured data points and the parameterization to 
extrapolate to $x = 0$.
The elastic contribution has not been excluded in the integral.
The results  for $\Gamma^p_1(Q^2)$ of the proton are shown in Figure \ref{fig:gammaboth}. 
The characteristic feature is the strong $Q^2$ dependence for $Q^2 < 1$ GeV$^2$, 
with a zero crossing near $Q^2=0.25$ GeV$^2$. The zero crossing is largely due to 
an interplay between the excitation strength of the $\Delta(1232)$ and the $S_{11}(1535)$,
and the rapid change in the helicity structure of the $D_{13}(1520)$ from helicity $3\over 2$ 
dominance at the real photon point to helicity $1\over 2$ dominance at $Q^2 > 0.5$ GeV$^2$. The 
latter behavior is well understood in dynamical quark models \cite{close}. A similar 
helicity flip is also observed for the $F_{15}(1680)$.

\begin{figure}[tb]
\vspace{90mm} 
\centering{\includegraphics{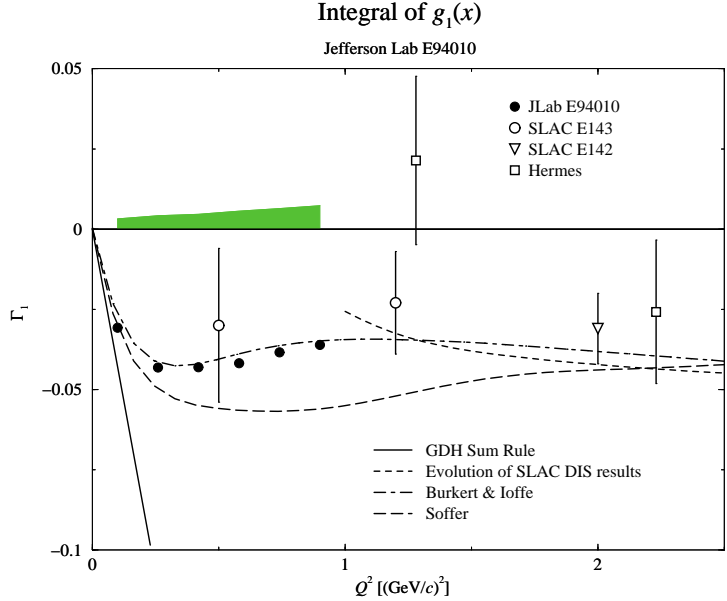}}
\caption{\small Preliminary results of the first moment of spin structure function $g_1$ for neutrons, 
corrected for the unmeasured deep inelastic part. The 
band indicates the size of systematic errors.}
\label{fig:gamma1n}
\end{figure}

Measurements on $ND_3$ have been carried out in CLAS \cite{kuhn}, 
and on  $^3He$ in Hall A \cite{halla1} to measure the 
corresponding integrals for the neutron. Here I only discuss the Hall A results.
Data were taken with  
the JLab Hall A spectrometers using a polarized $^3He$ target. Since the 
data were taken at fixed scattering angle, $Q^2$ and $\nu$ are correlated.
Cross sections at fixed $Q^2$ are determined by an interpolation between 
measurements at different beam energies. Both longitudinal and transverse 
settings of the target polarization were used. After correcting for nuclear effects and 
accounting for the deep inelastic part of the integral, the 
first moment of $g_1(x,Q^2)$ for neutrons can be extracted, and 
is shown in Figure \ref{fig:gamma1n}. 
The data deviate from the trend seen for the pQCD-evolved asymptotic 
behavior for $Q^2 < 1$ GeV$^2$. This is largely due to the 
contribution of the $\Delta(1232)$. The data are well described by a
model\cite{buriof} that includes resonance excitations and 
describes the connection to the deep inelastic regime assuming 
vector meson dominance. Another parametrization of the $Q^2$ dependence 
is from Soffer and Teryaev \cite{soffer2}.   

%%%%%%%%%%%%%%%%%%%%%%%%%%%%%%%%%%%%%%%%%%%%%%%%%%%
%%%%%%%%%%%%%%%%%%%%%%%%%%%%%%%%%%%%%%%%%%%%%%%%%%%%%

\section{The Bjorken Integral} 

\begin{figure}[tb]
\vspace{85mm} 
\centering{\includegraphics{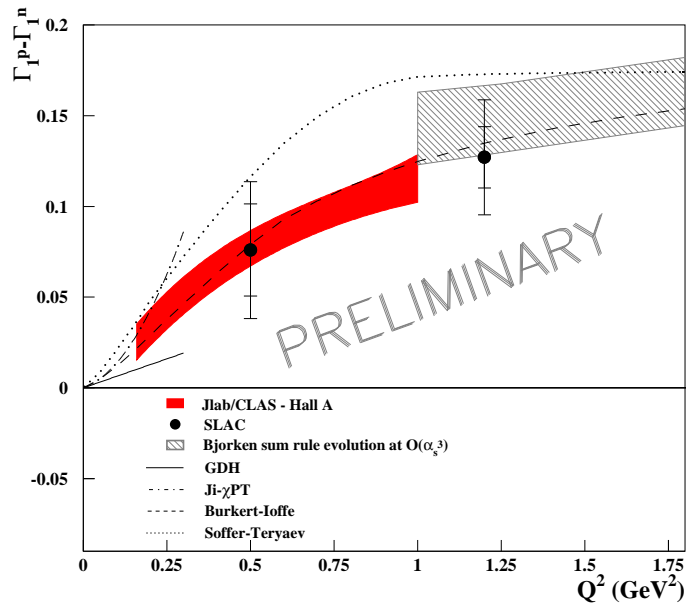}}
\caption{\small Preliminary results for the Bjorken integral 
of the proton-neutron difference. The band below $Q^2 = 1$ GeV$^2$ 
parametrizes the data and error for both data sets.}
\label{fig:gamma1pn}
\end{figure}

Using the results on $\Gamma_1(Q^2)$ for protons and neutrons 
one can determine the $Q^2$ dependence of the Bjorken integral 
$\Gamma_1^{(p-n)} = \Gamma_1^p - \Gamma_1^n$.
In this integral, all contributions of isospin $3\over 2$ resonances, 
such as the $\Delta(1232)$, drop out, and contributions of 
other resonances may be reduced as well. 
Also, since the GDH sum rule for the proton-neutron difference 
is positive, no zero-crossing is necessary to connect to  
the asymptotic behavior. The preliminary data are shown in 
Figure \ref{fig:gamma1pn}. Since the CLAS data and the Hall A 
data were measured at somewhat different $Q^2$ values, 
the data in each set were connected with a smooth interpolating curve and 
then subtracted. The resulting curve is the centroid of the 
shaded error band. The band at higher $Q^2$ corresponds to
the $O(\alpha_s^3)$ evolution of the Bjorken sum rule. At low $Q^2$ the
HBChPT curve seems to describe the trend of the data up 
to $Q^2 \approx 0.2$ GeV$^2$. A recent ChPT calculation\cite{meissner} 
in $O(p^4)$ predicts values significantly above the HBChPT curve. 
The model with explicit resonance contributions
gives a good description of the global behavior for both proton and neutron
targets, and for their difference.  
%%%%%%%%%%%%%%%%%%%%%%%%%%%%%%%%%%%%%%%%%%%%%%%%%%%
%%%%%%%%%%%%%%%%%%%%%%%%%%%%%%%%%%%%%%%%%%%%%%%%%%%%%

\section{Conclusions}

First high precision measurements of double polarization responses have been 
carried out at Jefferson Lab in a range of $Q^2$ not covered in previous 
high energy experiments. Spin structure functions and spin integrals $\Gamma_1(Q^2)$ 
have been extracted for protons and neutrons. 
The proton and neutron data both show large contributions from resonance 
excitations. $\Gamma_1^p(Q^2)$ shows a dramatic change with $Q^2$, 
including a sign change near $Q^2 = 0.25$ GeV$^2$, while $\Gamma_1^n(Q^2)$ remains
negative, however is strongly affected by the $\Delta(1232)$ contribution. 
Qualitatively, the strong deviations from the trend of the deep inelastic behavior 
for $Q^2 < 1$ GeV$^2$ mark the transition from the domain of single and multiple 
parton physics to the domain of resonance excitations 
and hadronic degrees of freedom. 

New data have been taken both on hydrogen and deuterium with nearly 10 times
more statistics, and higher target polarizations, and cover a larger range 
of energies from 1.6 GeV to 5.75 GeV. The year 2001 data cover a  
$Q^2$ range from 0.05 to 3 GeV$^2$, and a larger
part of the deep 
inelastic regime than the data presented here. 
This will allow a reduction of the systematic uncertainties
related to the extrapolation to $x = 0$. Moreover, since data are available 
at fixed $Q^2$ taken at different beam energies, a separation of $A_1(Q^2, W)$ and 
$A_2(Q^2, W)$ will be possible. The new data will also 
give much 
better sensitivity to resonance production in exclusive channels, such as 
$ep\rightarrow en\pi^+$, that have been measured previously\cite{devita}. 
Finally, at the higher energies, CLAS will be able to 
study single and double spin asymmetries in various exclusive and 
semi-inclusive reactions currently of great interest to access the 
transverse quark distribution functions\cite{harut}.

There is a program underway in JLab Hall A to measure the GDH 
integral for neutrons down to extremely small $Q^2$ values\cite{halla2}, 
near the real photon point, and to measure the asymmetry $A_1(x,Q^2)$ 
for the neutron at high $x$~\cite{halla3}. High precision data for $A_1$ and $A_2$ at 
$Q^2 = 1$ GeV$^2$ are also expected to come from experiment E-01-006 
in Hall C\cite{rondon}.

\vspace{0.5cm}\noindent
The Southeastern University Research Association (SURA) operates JLab for the 
U.S. Department of Energy under Contract No. DE-AC05-84ER40150.

%\section*{References}


\begin{thebibliography}{200} 
\bibitem{ashman} J. Ashman et al., Nucl. Phys. B328, 1 (1989) 
\bibitem{filipone} for a recent overview see: B. Filipone and X. Ji,
 Adv. Nucl. Phys. 26, 1 (2001).  
\bibitem{bjorken} J.D. Bjorken, Phys. Rev. 179, 1547 (1969)
\bibitem{gerasimov} S.B. Gerasimov; Sov. J. Nucl. Phys. 2, 430 (1966).  
\bibitem{drell}S.D. Drell and A.C. Hearn, Phys. Rev. Lett.16, 908 (1966).
\bibitem{ahrens} J. Ahrens et al., Phys.Rev.Lett.87:022003 (2001); also: talk at GDH2002.
\bibitem{jios} X. Ji, J. Osborne, J. Phys. G27 127 (2001). 
\bibitem{ji} X. Ji, in: Excited Nucleons and Hadronic Structure, World Scientific (2001) 
\bibitem{alexandrou} C. Alexandrou et al., hep-lat/0209074 (2002).
\bibitem{abe} K. Abe et al., Phys. Rev. D58, 2003 (1998).
\bibitem{burli} V. Burkert and Zh. Li, Phys. Rev. D47, 46 (1993).
\bibitem{buriof} V. Burkert, B. Ioffe, Phys. Lett. B296, 223 (1992); 
J. Exp. Theo. Phys.78, 619(1994).  
\bibitem{minehart} R. Minehart, in: NSTAR2001, World Scientific, 
eds. D. Drechsel, L. Tiator (2001)
\bibitem{bloom} E.D. Bloom and F.G. Gilman, Phys, Rev. Lett. 25, 1140 (1970)
\bibitem{cloisg} F.E. Close and N. Isgur, Phys. Lett. B509, 81 (2001)
\bibitem{burkert_elba} For a recent overview see: V.D. Burkert, hep-ph/0210321
\bibitem{soffer} J. Soffer and O. Teryaev, Phys. Rev. D51, 25 (1993)
\bibitem{xji} X. Ji, C.W. Kao, J. Osborne, Phys.Lett.B472:1-4 (2000) 
\bibitem{burk} V. Burkert, Phys. Rev. D63, 97904 (2001)  
\bibitem{close} F. E. Close and F.J. Gilman, Phys. Letts. 38B, 541 (1972).   
\bibitem{bucramin} V. Burkert, D. Crabb, R. Minehart, et al., 
JLab experiment E-91-023.
\bibitem{kuhn} S. Kuhn, G. Dodge, M. Taiuti, et al., JLab experiment E-93-009.
\bibitem{devita} R. De Vita et al., Phys. Rev. Letts. 88, 082001-1 (2002).
\bibitem{harut} H. Avakian, talk at SPIN2002, to be published in the proceedings.
\bibitem{halla1} G. Cates, J.P. Chen, Z.E. Meziani et al., JLab experiment E-94-010. 
\bibitem{soffer2} J. Soffer and O.V. Teryaev, Phys. Rev.D56, 7458 (1997). 
\bibitem{meissner} V. Bernard, T. Hemmert, and U. Meissner, Phys. Lett. B545, 105-111, (2002).  
\bibitem{halla2} J.P. Chen, A. Deur, and F. Garibaldi, et al., Jlab Experiment E-97-110.
\bibitem{halla3} J.P. Chen, Z.E. Meziani, P. Souder, et al., Jlab Experiment E-99-117.
\bibitem{rondon} O. Rondon, et al., Experiment E-01-006 (2001). 

\end{thebibliography}
\end{document}